\documentclass[final,5p,times,twocolumn]{elsarticle}

\usepackage{graphicx}

\usepackage{amssymb}

\journal{Nuclear Instruments and Methods A}

\begin{document}

\begin{frontmatter}



\title{Performance of Drift-Tube Detectors at High Counting Rates for High-Luminosity LHC Upgrades}


\author[A]{Bernhard Bittner}
\author[A]{J\"org Dubbert\fnref{label1}} 
\author[A]{Oliver Kortner}
\author[A]{Hubert Kroha\corref{label2}}
\author[A]{Alessandro Manfredini}
\author[A]{Sebastian Nowak} 
\author[A]{Sebastian Ott}
\author[A]{Robert Richter}
\author[A]{Philipp Schwegler}
\author[A]{Daniele Zanzi}
\author[B]{Otmar Biebel} 
\author[B]{Ralf Hertenberger}
\author[B]{Alexander Ruschke}
\author[B]{Andre Zibell}
\address[A]{Max-Planck-Institut f\"ur Physik, Munich, Germany}
\address[B]{Ludwig-Maximilians University, Garching, Germany}

\fntext[label1]{Now at University of Michigan.}
\cortext[label2]{Corresponding author.}
\emailauthor{kroha@mppmu.mpg.de}{Hubert Kroha}

\begin{abstract}

The performance of pressurized drift-tube detectors at very high background 
rates has been studied at the Gamma Irradiation Facility (GIF) at CERN 
and in an intense 20~MeV proton beam at the Munich Van-der-Graaf tandem accelerator 
for applications in large-area precision muon tracking at high-luminosity 
upgrades of the Large Hadron Collider (LHC). The ATLAS muon drift-tube (MDT) 
chambers with 30~mm tube diameter have been designed to cope with $\gamma$ and neutron 
background hit rates of up to 500~Hz/cm$^2$. Background rates of up to 14~kHz/cm$^2$ 
are expected at LHC upgrades. The test results with standard MDT readout electronics 
show that the reduction of the drift-tube diameter to 15~mm, while leaving the 
operating parameters unchanged, vastly increases the rate capability well beyond 
the requirements. The development of new small-diameter muon drift-tube (sMDT) chambers
for LHC upgrades is completed. Further improvements of tracking efficiency and spatial resolution
at high counting rates will be achieved with upgraded readout electronics employing 
improved signal shaping for high counting rates.

\end{abstract}


\end{frontmatter}



\section{Introduction}

The muon systems of the LHC experiments require high precision tracking
detectors covering very large areas. Drift tube detectors provide a robust and
efficient solution even at the high background rates of neutrons and $\gamma$ rays 
experienced in the muon spectrometer of the ATLAS experiment
which is equipped with muon drift-tube (MDT) chambers~\cite{MDT1,MDT2} with 30~mm tube diameter.
For the high-luminosity upgrades of the LHC, the background hit rate is expected to
increase by almost an order of magnitude compared to the LHC design luminosity, reaching
a maximum of about 14~kHz/cm$^2$ in the forward regions of the ATLAS muon spectrometer. 
Besides requirements for higher selectivity of high momentum tracks and, therefore, 
improved momentum resolution at the first level
of the muon trigger~\cite{trigger}, muon tracking chambers with increased rate capability are
needed in the high-radiation regions of the muon detectors. For the muon tracking chambers 
of the ATLAS experiment, which are operated in the field of superconducting air-core toroid
magnets, a track segment resolution per chamber of better than $60~\mu$m and, correspondingly, 
an anode wire positioning accuracy of better than $20~\mu$m is required.

\begin{figure}[htb] 
\centering

\includegraphics[width=0.8\columnwidth,keepaspectratio]{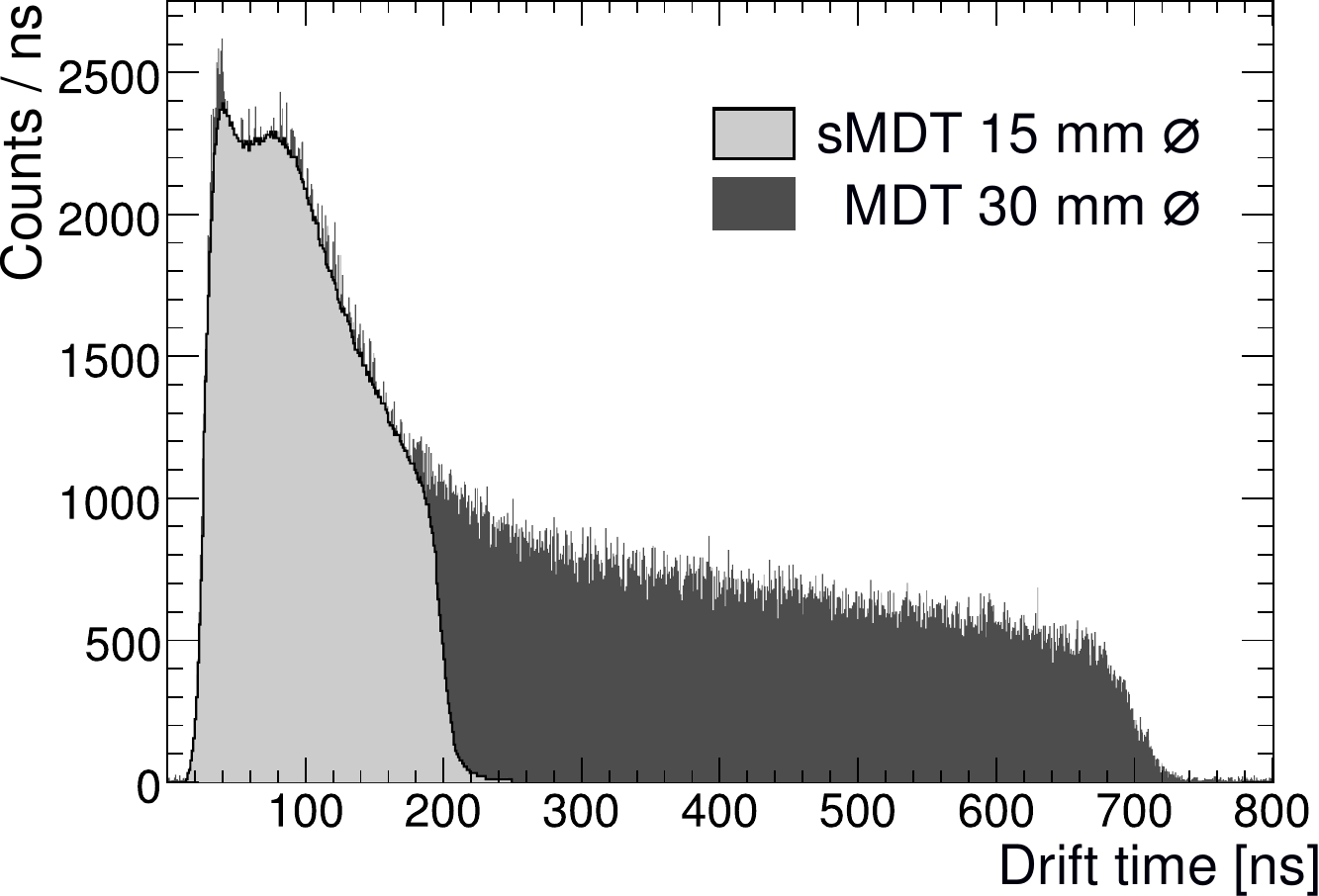} 
\vspace{-2mm}

\caption{The measured drift time spectra of 30 and 15~mm diameter drift tubes with maximum drift times
of about 700~ns and 185~ns, respectively.} 
\label{fig:drift_time_spectra}
\end{figure}

\begin{figure}[htb] 
\centering

\includegraphics[width=0.75\columnwidth,keepaspectratio]{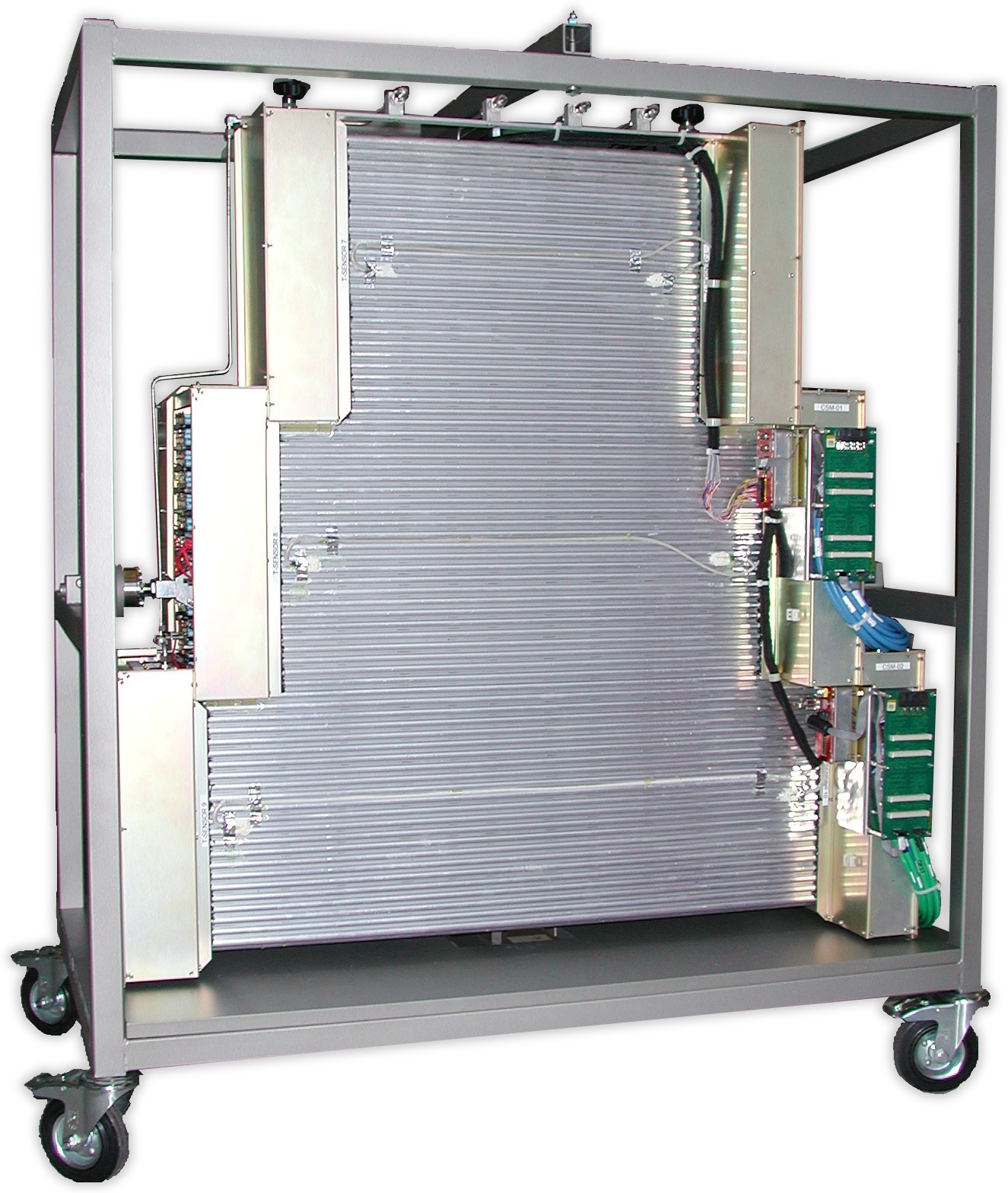}
\vspace{-2mm}

\caption{The sMDT prototype chamber of a size of one square meter.} 
\label{fig:prototype}

\centering
\includegraphics[width=0.95\columnwidth,keepaspectratio]{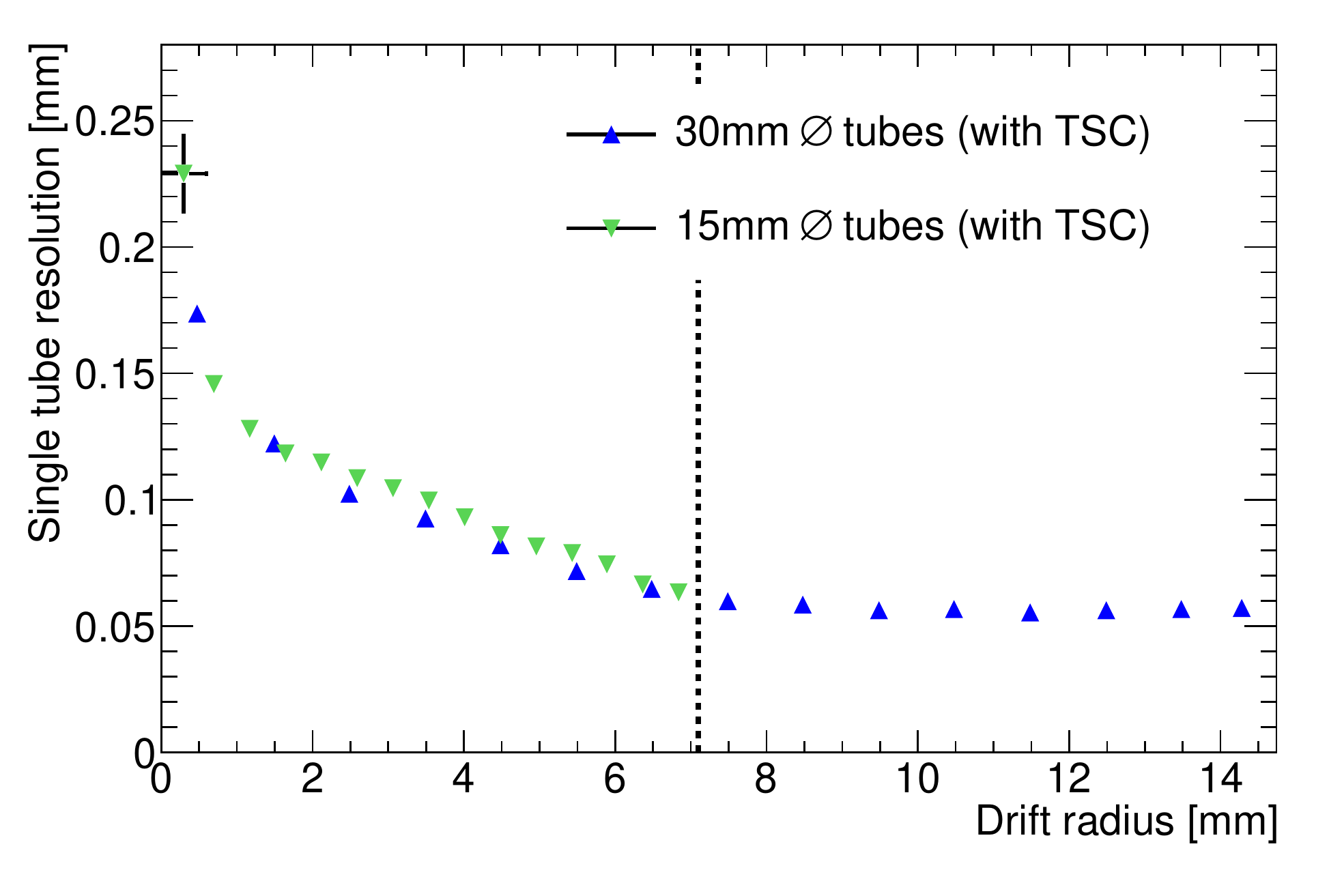}
\vspace{-5mm}

\caption{Measurement of the spatial resolution of 15 compared to 30~mm diameter drift tubes
as a function of the drift radius r in high-energy muon beams at CERN without background irradiation. 
Time slewing corrections (TSC) have been applied using the pulse height information of the readout electronics. 
The resolution curves overlap in the common radial region as expected.} 
\label{fig:resolution_r}
\end{figure}

\begin{figure}[htb] 
\centering

\includegraphics[width=0.8\columnwidth]{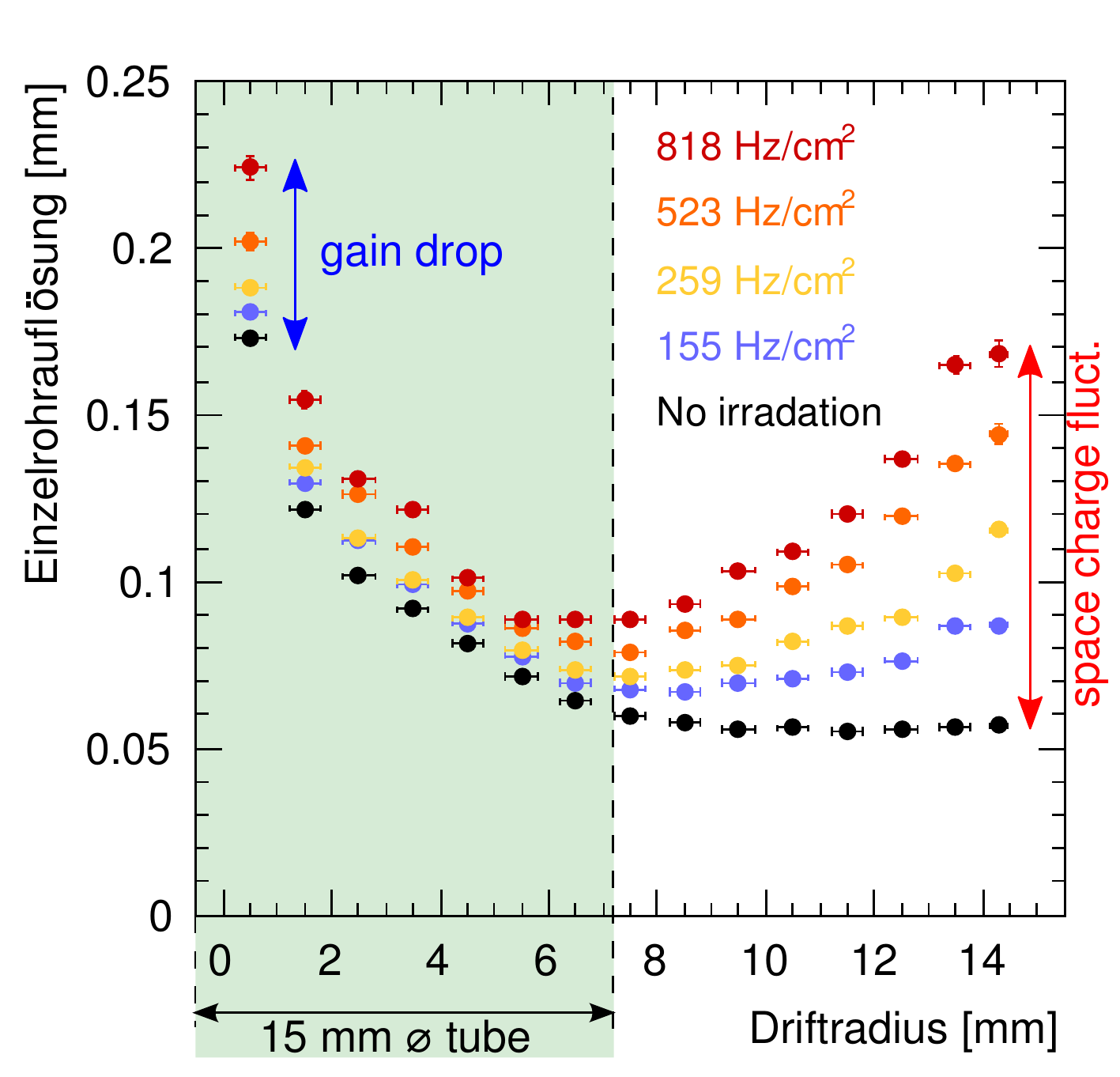} 
\vspace{-2mm}

\caption{Spatial resolution of 30~mm diameter drift tubes as a function of the drift radius r
measured for increasing $\gamma$ hit rate at the CERN Gamma Irradiation facility~\cite{GIF2004}
(dotted curves from bottom to top in the same order as the corresponding background rates in the legend).} 
\label{fig:resolution_vs_rate_30mm}

\centering 

\includegraphics[width=\columnwidth,clip]{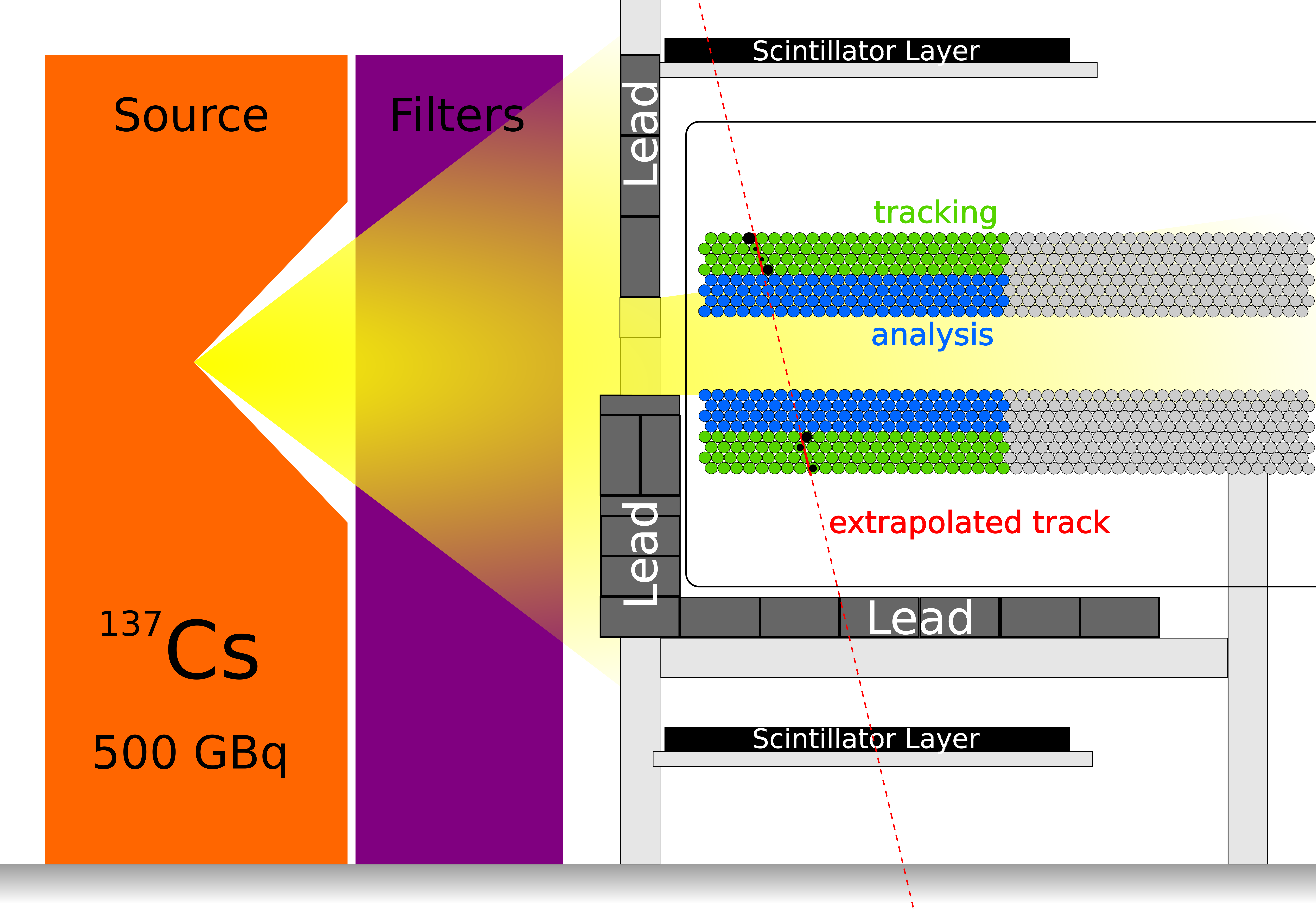} 
\vspace{-3mm}

\caption{The test setup of the sMDT prototype chamber in the CERN Gamma Irradiation Facility. 
The unirradiated regions of the chamber serve as tracking reference for cosmic ray muons.} 
\label{fig:GIF_setup}
\end{figure}

\begin{figure}[!htb] 

\includegraphics[width=0.63\columnwidth,keepaspectratio]{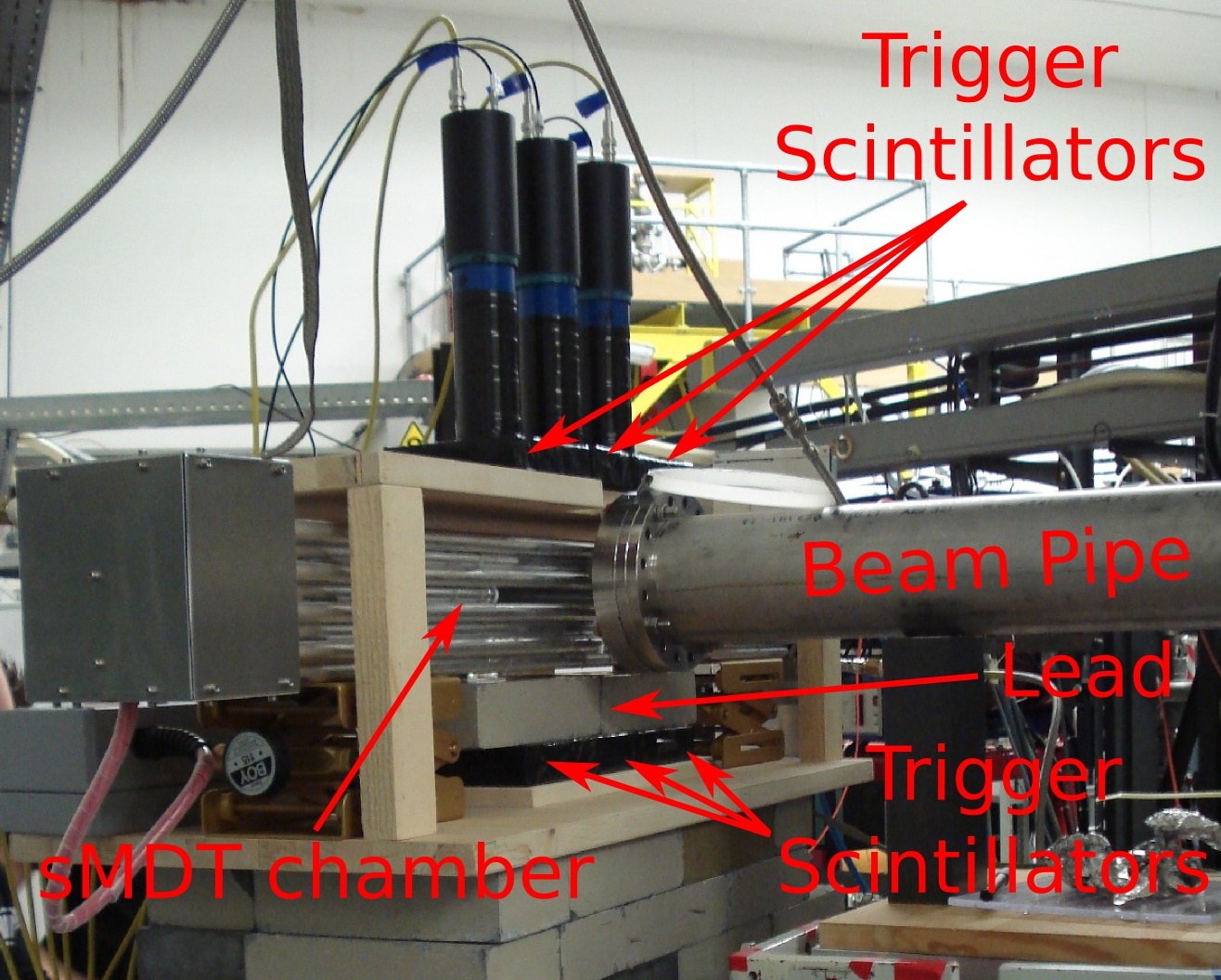}
\includegraphics[width=0.35\columnwidth,clip]{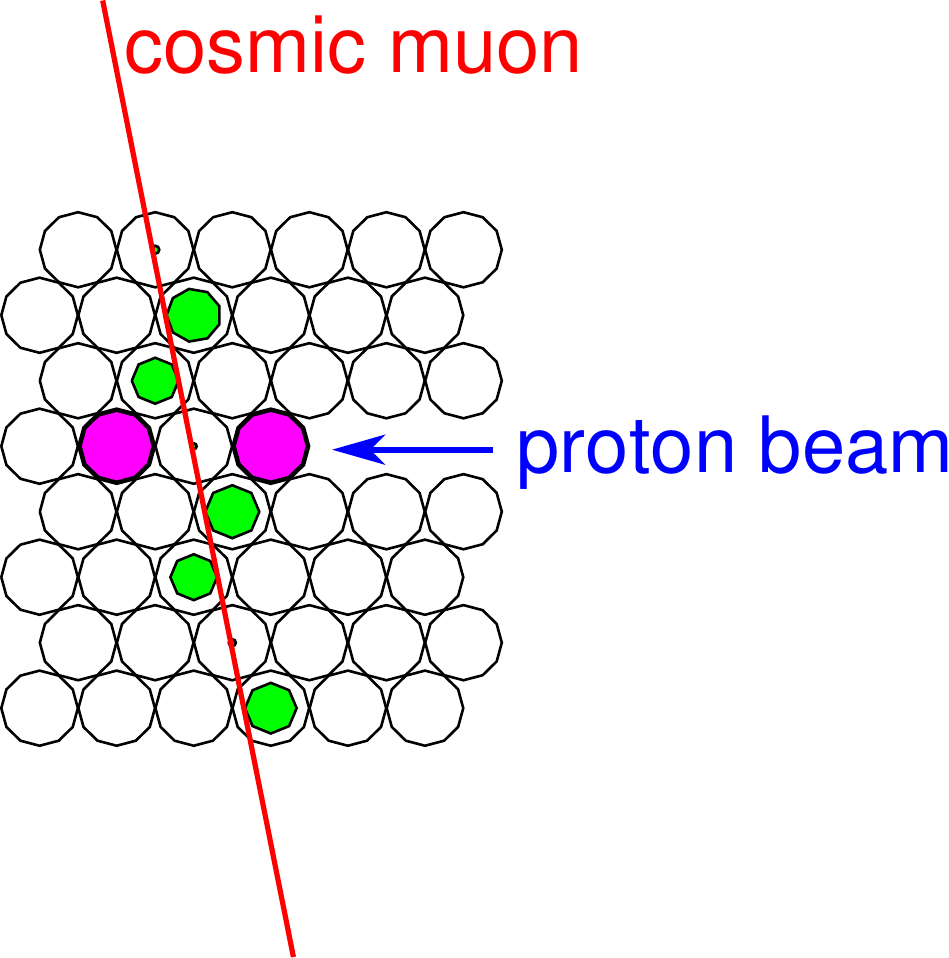}

\caption{Test setup of a sMDT chamber in a 20\,MeV proton beam delivered by the 
        Munich Van-der-Graaf Tandem accelerator. The beam irradiates    
        only the tubes of the middle layer (right, filled circles) over a longitudinal section of about 7~cm. 
	The other tube layers serve as tracking reference 
        for cosmic ray muons. The trigger segmentation along the tubes allows for the 
        selection of muon tracks passing or missing the illuminated region.} 

\label{fig:tandem_setup}

\centering
\includegraphics[width=0.8\columnwidth,keepaspectratio]{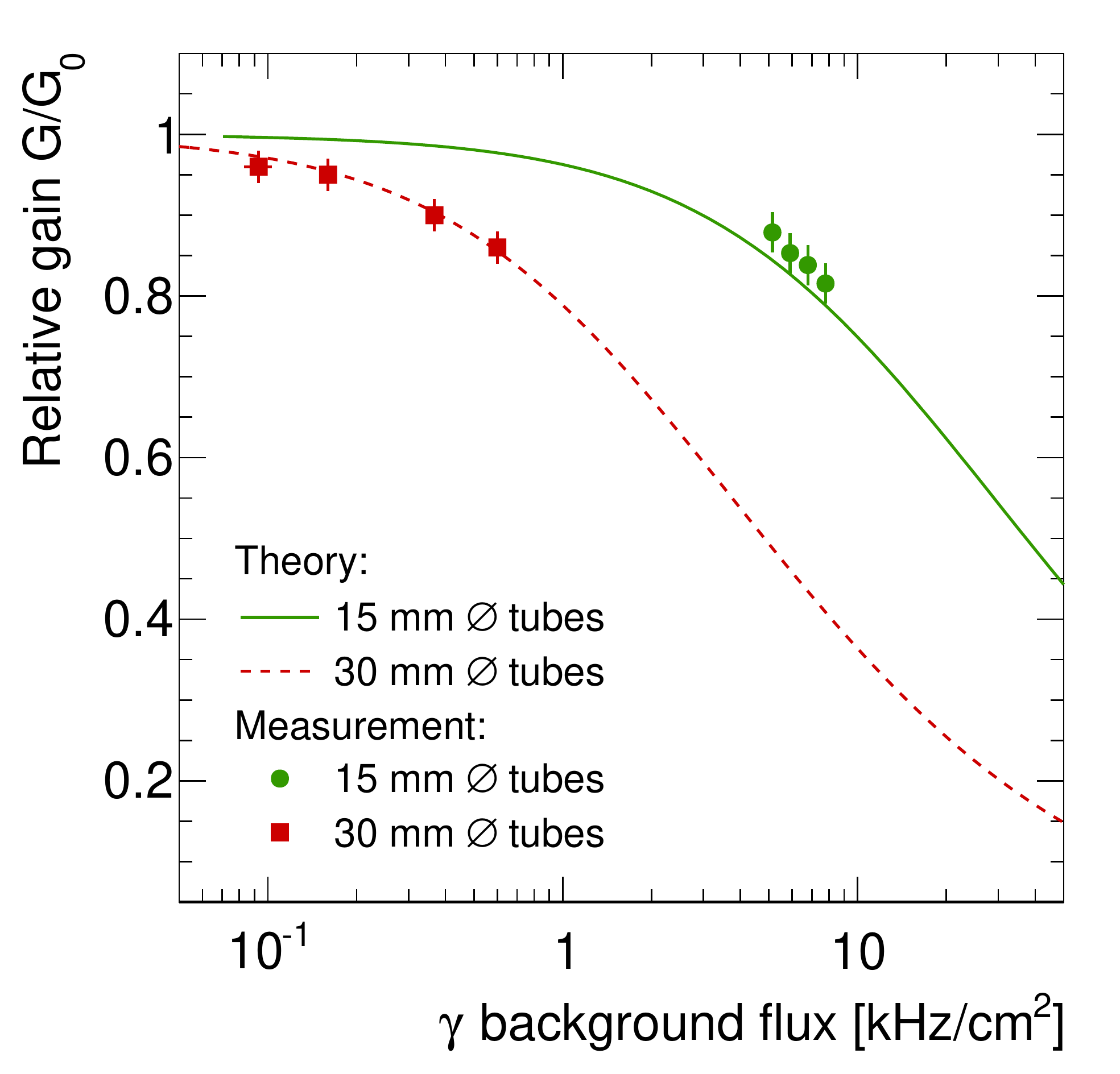}
\vspace{-5mm}

\caption{Measurements of the gas gain normalised to the nominal gain G$_0$ as a function of the 
$\gamma$ background hit rate at GIF for 15 and 30\,mm diameter drift tubes 
compared to predictions based on the Diethorn model~\cite{Diethorn} and taking into account with increasing background 
flux also saturation of the space charge production due to reduction of the electric field at the sense wire.} 

\label{fig:gaindrop_gif}
\end{figure}

For these purposes, new small-diameter muon drift-tube (sMDT) chambers with 15 instead of 30~mm outer tube diameter 
have been developed. They are operated with the same sense wire diameter of $50~\mu$m, the same
gas mixture of Ar:CO$_2$ (93:7) at 3 bar absolute pressure, and the same nominal gas gain of G$_0=20000$ 
as the ATLAS MDT chambers. These parameters correspond to an operating voltage of 2730~V and a maximum
drift time of 185~ns which is by about a factor of 3.8 shorter than for the 30~mm diameter as shown in 
Fig.~\ref{fig:drift_time_spectra}. Together with the twice smaller tube cross section, 
this leads to 7.6 times smaller occupancy per unit tube length and to a vast improvement of the 
tracking efficiency and also of the spatial resolution at high rates.

\section{sMDT chamber construction}

A complete sMDT prototype chamber~\cite{prototype} has been constructed with 8 layers of up to 1~m long 
aluminum tubes of 0.4~mm wall thickness mounted on either side of an aluminum space frame 
(see Fig.~\ref{fig:prototype}). Using precise mechanical jigging, each of the two multilayers of tubes has been assembled within one working day.
The anode wire positions have been measured with a resolution of about $5~\mu$m 
with cosmic ray tracks using two large MDT chambers with precisely known wire positions as tracking reference. 
The wire positioning accuracy has been measured to be better than $16~\mu$m.

\section{High-rate performance}

The spatial resolution as a function of the drift radius of the 15~mm diameter drift tubes 
of the sMDT prototype chamber has been measured in the absence of background radiation in a high-energy 
muon beam at CERN (see Fig.~\ref{fig:resolution_r}). In the common radial range, it agrees
very well with the resolution measured for 30~mm diameter MDT tubes under the same conditions~\cite{GIF2004}.
Without background radiation, the spatial resolution averaged over the tube radius is worse for smaller-diameter tubes
since the resolution improves with increasing drift distance. After time slewing corrections
using the pulse height information provided by the readout electronics, the average resolution of
15~mm diameter tubes is $106\pm 2~\mu$m compared to $83\pm 2~\mu$m for the 30~mm diameter tubes~\cite{GIF2004}.

With increasing flux of ionizing background radiation, the resolution deteriorates with increasing
drift radius due to fluctuations in the space-charge density of the positive ions, and  
also at small radii due to the gas gain reduction caused by the shielding of the wire potential 
by the space charge. Fig.~\ref{fig:resolution_vs_rate_30mm} shows that the effect of space charge
fluctuations dominating for 30~mm diameter tubes in the radial range above 7.5~mm is strongly suppressed for 15~mm tube diameter.
The gain loss at given primary ionisation from background radiation 
is to first approximation proportional to the inner tube radius to the third power 
and, therefore, is smaller by a factor of $(14.6~mm/7.1~mm)^3=8.7$ for 15 compared to 30~mm diameter tubes.  
The dependence of the gas gain on the $\gamma$ background flux from a 500~GBq ${}^{137}$~Cs source has been measured 
for 30 and 15~mm diameter tubes in the Gamma Irradiation Facility at CERN
using the pulse height information of the readout electronics
(see Fig.~\ref{fig:gaindrop_gif}). It is well described by models obtained by iterating the Diethorn prediction~\cite{Diethorn} 
for the gas gain as a function of the electric field at the sense wire with the space charge production in this field and shows the
expected suppression of the gain loss in 15~mm diameter tubes.

The relevant quantity for evaluating the chamber tracking efficiency and resolution in the presence of background 
is the so called $3\sigma$ single-tube efficiency which is defined as the probability of a muon hit in a drift tube to lie
on the reconstructed muon track within three times the drift tube resolution $\sigma$. 
Background signals may mask subsequent muon hits for the duration
of the signal pulse length and the electronics deadtime leading to a decrease of the $3\sigma$ efficiency with increasing
background counting rate. The standard ATLAS MDT readout electronics~\cite{electronics}
used in all performance tests allows for an adjustment of the deadtime between a minimum value of 175~ns
and the maximum of 790~ns which is used for the ATLAS MDT chambers in order to suppress
secondary hits from late arriving ionization clusters. With the minimum deadtime setting which 
is applicable for the much shorter drift times of the sMDT tubes, the degradation rate
of the $3\sigma$ efficiency is expected to improve by about a factor of 9 compared to 30~mm diameter tubes 
where a factor of 2 is due to the twice smaller tube cross section exposed to the radiation.

\begin{figure}[h!]
\centering

\includegraphics[width=\columnwidth,keepaspectratio]{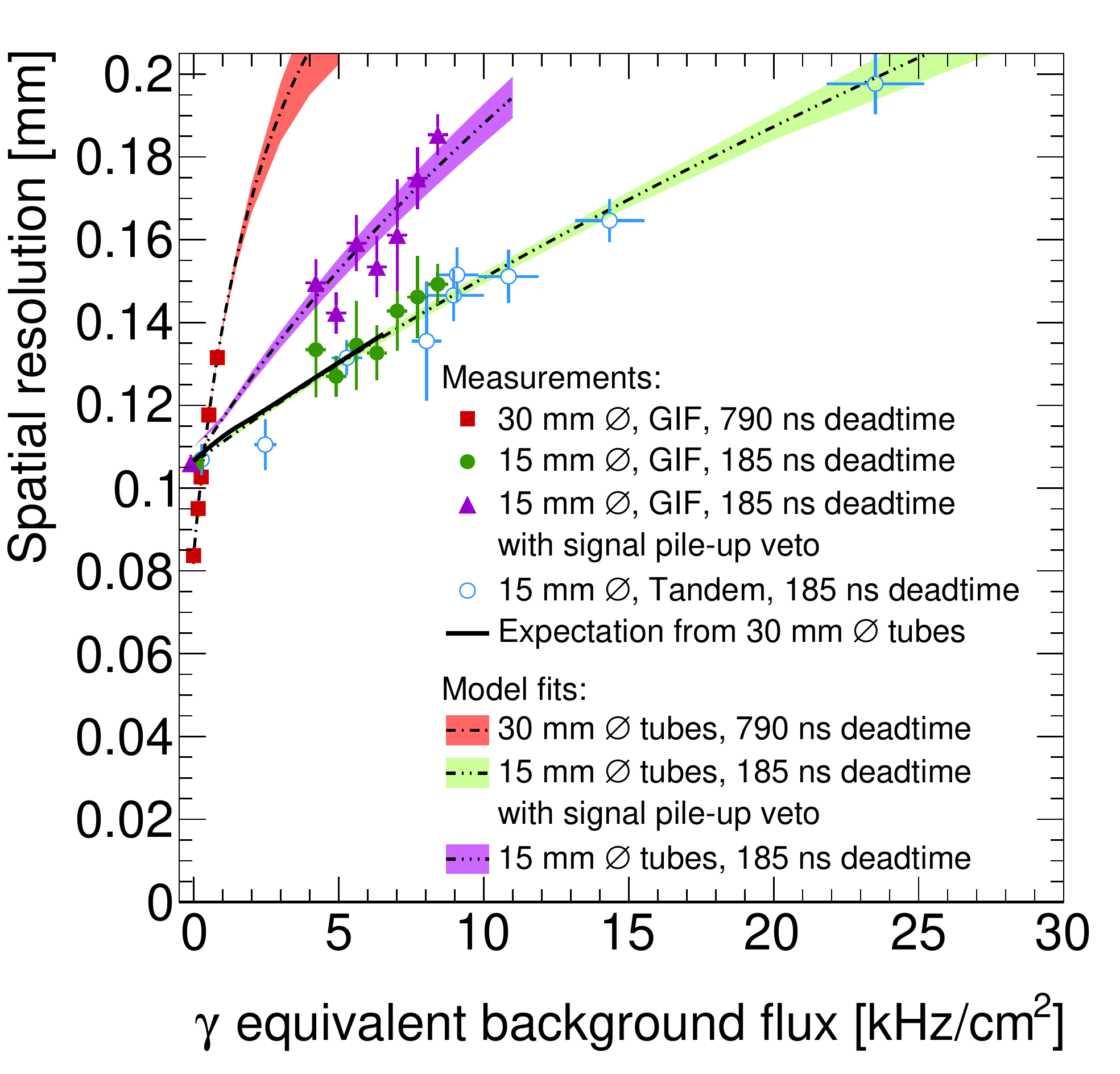}
\vspace{-9mm}

\caption{Average spatial resolution of  
15~mm diameter drift tubes with and without electronic signal pile-up effect
under $\gamma$ (GIF) and proton (Tandem) irradiation 
as a function of the equivalent $\gamma$ hit rate (see text) compared to the
results for 30~mm diameter tubes.
The maximum background rate expected in the ATLAS muon detector 
at high-luminosity LHC is 14~kHz/cm$^2$.}
\label{fig:resolution_n}    

\centering
\includegraphics[width=\columnwidth,keepaspectratio]{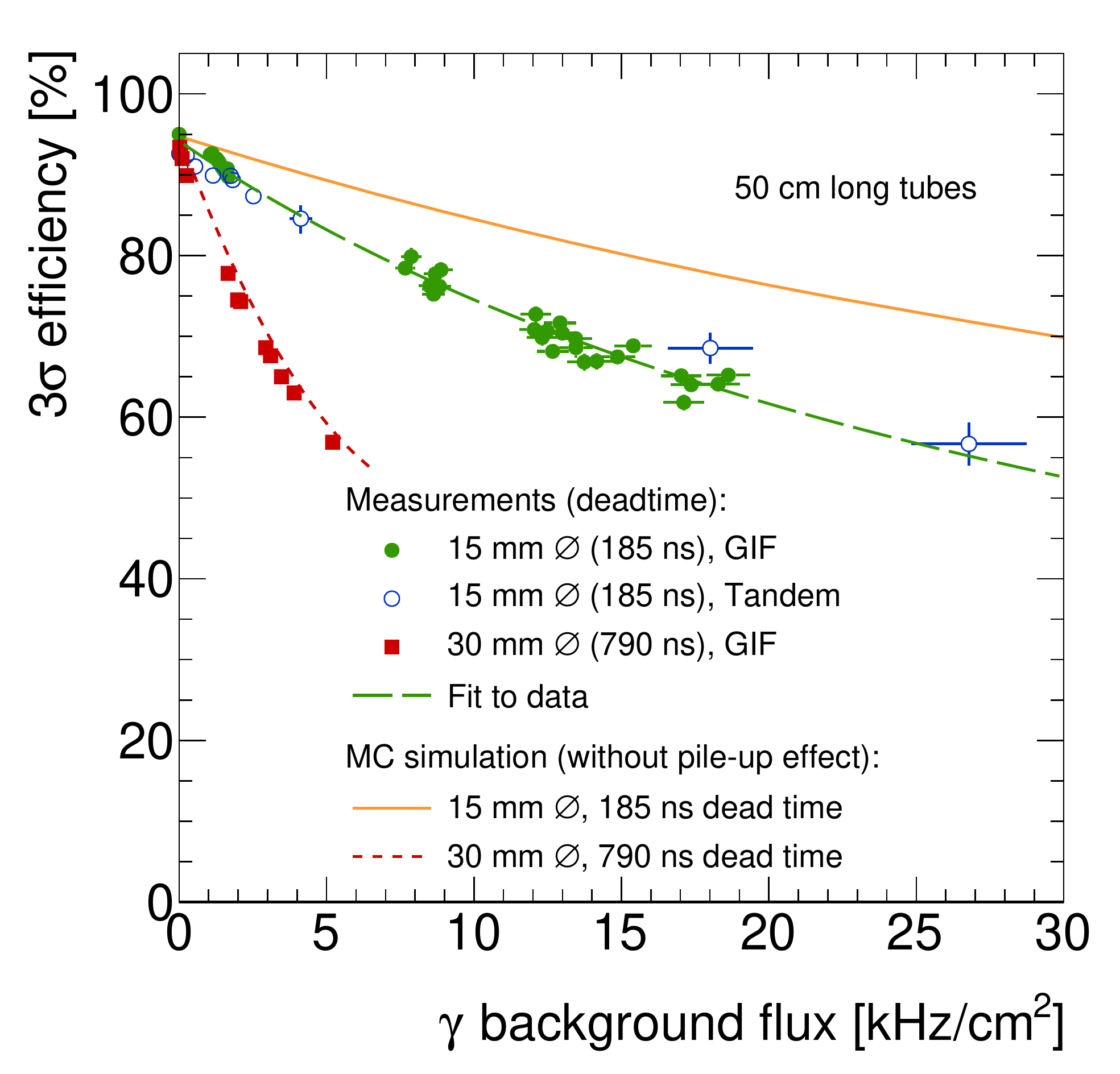}
\vspace{-9mm}

\caption{Dependence of the $3\sigma$ tracking efficiency of
15~mm diameter drift tubes on the background counting rate
expressed in terms of the corresponding uniform $\gamma$ hit rate in 0.5~m long tubes
compared to the results for 30~mm diameter tubes and to the expectation 
without electronic signal pile-up effect. Without background irradiation
the $3\sigma$ efficiency is $94\%$ due to $\delta$-rays 
created by muon interactions in the tube walls which produce signals earlier than the muons.
The maximum background rate expected in the ATLAS muon spectrometer
at high-luminosity LHC is 14~kHz/cm$^2$.}
\label{fig:efficiency_n}     
\end{figure}

\begin{figure}[htb]
\centering
\includegraphics[width=1.1\columnwidth,keepaspectratio]{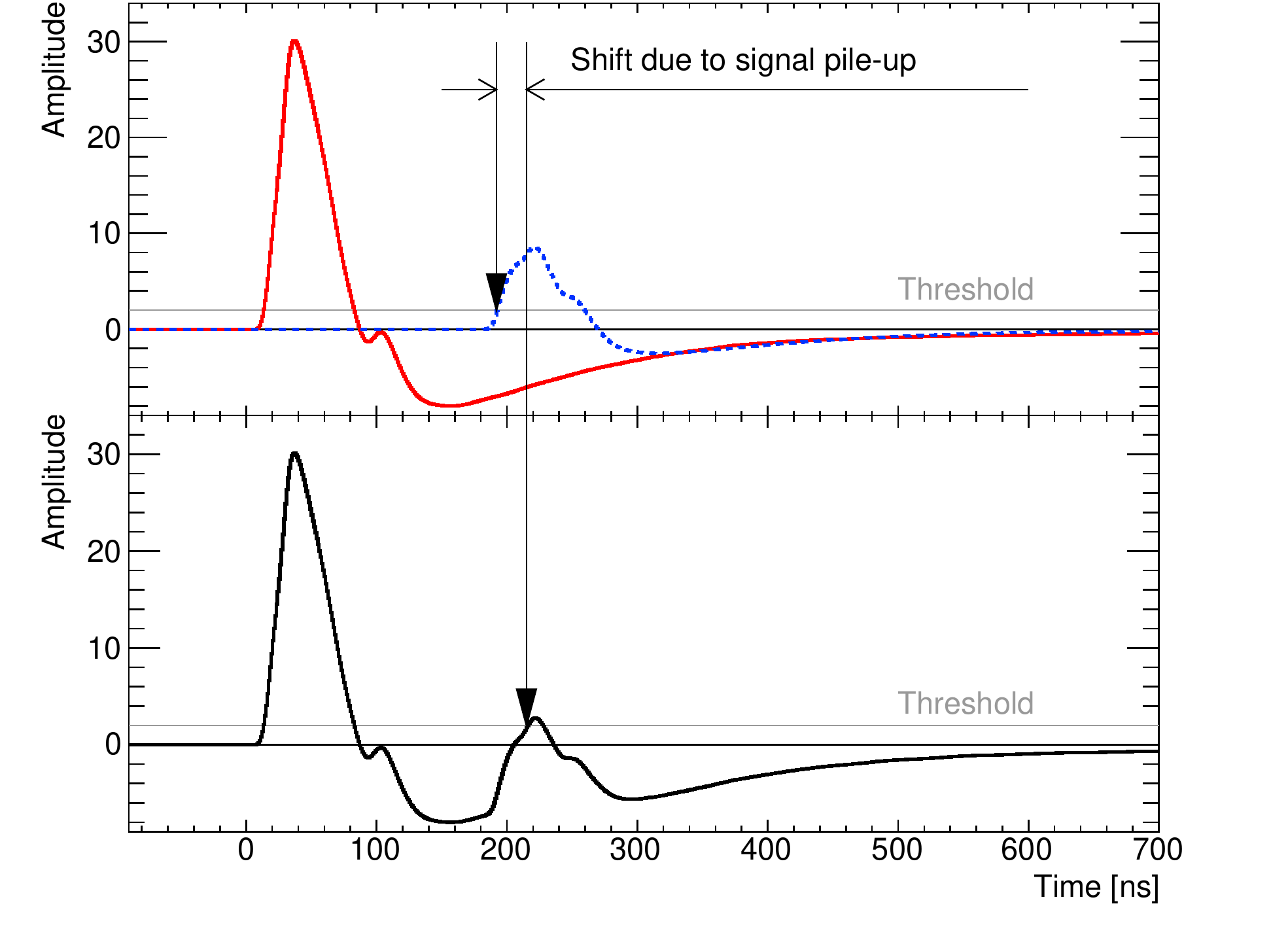}
\vspace{-9mm}
    
\caption{Illustration of the signal pile-up effects present for the standard MDT readout electronics
with bipolar shaping and minimum deadtime setting when used for the 15~mm diameter drift tubes. 
The red and blue curves in the top part of the graph show a background $\gamma$ or neutron pulse followed by a muon pulse 
independent of each other, while the black curve in the bottom part shows the superposition
of the two pulses. The large undershoots below the baseline of large background pulses can lead to the 
loss of subsequent muon hits and, in general, cause additional time slewing.}
\label{fig:signal_pileup}    
\end{figure}

The high-rate performance of the sMDT chambers equipped with standard MDT readout electronics has been studied with cosmic ray tracks 
in the Gamma Irradiation Facility (GIF) at CERN (see Fig.~\ref{fig:GIF_setup}) at $\gamma$ hit rates of up to
8.5~kHz/cm$^2$ corresponding to counting rates of up to 1200~kHz/tube and in a 20~MeV
proton beam at the Munich Van-der-Graaf Tandem accelerator (see Fig.~\ref{fig:tandem_setup}) at counting rates of up to 
1400~kHz/tube. In contrast to the uniform illumination of the tubes at GIF, the proton irradiation is localised to 
a 7~cm wide longitudinal section of one tube. The trigger segmentation allows for the selection of muon tracks traversing
either the irradiated or the unirradiated regions of the tube for the measurements of the resolution depending on the space charge
and of the efficiency depending on the counting rate, respectively.  
Equivalent $\gamma$ fluxes corresponding to the proton rates have been determined using 
the 15~mm diameter gain curve in Fig.~\ref{fig:gaindrop_gif} and measurements of the gas gain (see above) under proton irradiation.
The results for the average spatial resolution and the $3\sigma$ tracking efficiency 
in Figs.~\ref{fig:resolution_n} and \ref{fig:efficiency_n} show vast 
improvements from 30 to 15~mm tube diameter tubes increasing with the background rate. 
The $3\sigma$ efficiency measurements at GIF contain also hit efficiency loss due to gain loss which is, however,
negligible at the $\gamma$ rates reached at GIF.

With increasing counting rates, effects of the pile-up of muon signals on preceding background pulses
deteriorate the resolution and efficiency of the 15~mm diameter tubes operated with short deadtime (see Fig.~\ref{fig:signal_pileup}).
If a minimum time separation between successive hits of 600~ns is required (pile-up veto), the measured spatial resolution 
is considerably improved and agrees very well with the expectation 
from the 30~mm diameter tube measurements with maximum deadtime and no pile-up effects observed
(see Figs.~\ref{fig:resolution_n} and \ref{fig:efficiency_n})) which is
derived by averaging the data points in Fig.~\ref{fig:resolution_vs_rate_30mm}
only up to a radius of 7.1~mm and scaling the $\gamma$ hit rate with the gain ratio of 8.7. 
The $3\sigma$ efficiency of the 15~mm diameter tubes can also be considerably improved by suppressing the pile-up effect 
(see Fig.~\ref{fig:efficiency_n}). The pile-up effects will be eliminated in an upgraded version of the  
readout electronics which is under development by employing active baseline restoration.

From the results in Figs.~\ref{fig:resolution_n} and \ref{fig:efficiency_n}, one derives
a spatial resolution of better than $40~\mu$m and a tracking efficiency of better than $99\%$  
for 2 times 6 layer sMDT chambers at the maximum background rates expected in the ATLAS muon 
spectrometer at high-luminosity LHC.

\section{Ageing tests} 

Ar:CO$_2$ gas and only materials already certified for the ATLAS MDT chambers are used for the sMDT drift tubes in order to prevent ageing.
No ageing has been observed in MDT tubes up to a charge of 0.6~C/cm collected on the anode wire~\cite{MDT1}.
The sMDT tubes~\cite{prototype}, including the plastic material of the endplugs, have been irradiated with a 200 MBq ${}^{90}$Sr 
source over a period of 4 months accumulating an integrated charge of 6 C/cm on the wire without sign of ageing.

\section{Conclusions}

The development of fast high-precision muon drift-tube detectors (sMDT chambers) for high-luminosity LHC upgrades is completed. 
The high-rate performance of the new chambers exceeds the requirements for operation at the highest background rates at high-luminosity
LHC upgrades. Several sMDT chambers are under construction for the upgrade of the ATLAS
muon spectrometer in the 2013/14 LHC shutdown period. A smaller sMDT chamber has already
been operated in the highest irradiation region of the ATLAS muon spectrometer during 2012
data taking confirming the expected performance.




\end{document}